\documentclass[conference]{IEEEtran}
\IEEEoverridecommandlockouts
\usepackage{amsmath,amssymb,amsfonts,hyperref}
\usepackage{makecell}
\usepackage{graphicx}
\usepackage{subcaption}  
\usepackage{caption}
\usepackage{amssymb}
\usepackage{cite}
\usepackage{multirow}
\usepackage{colortbl,xcolor}
\usepackage{algorithmic}
\usepackage{textcomp}
\usepackage{booktabs}
\definecolor{blue}{RGB}{220,235,250}
\def\BibTeX{{\rm B\kern-.05em{\sc i\kern-.025em b}\kern-.08em
    T\kern-.1667em\lower.7ex\hbox{E}\kern-.125emX}}

\begin{document}

\title{LLMCodec: Adapting Video Codecs for Efficient Weight Compression of Large Language Models}

\author{
\IEEEauthorblockN{Rui Wang, Yan Zhao,  Li Song, Zhengxue Cheng\textsuperscript{*}}
\IEEEauthorblockA{\textit{Institute of Image Communication and Network Engineering, Shanghai Jiao Tong University.}\\
Shanghai, China\\
\{wangrui\_5056, zhaoyanzy, song\_li, 
zxcheng\}@sjtu.edu.cn}
\thanks{\textsuperscript{*}Corresponding author: Zhengxue Cheng (zxcheng@sjtu.edu.cn).}
}

\maketitle
\begin{abstract}

The rapid development of large language models (LLMs) has led to remarkable advances in natural language processing. However, the increasing scale of these models introduces substantial challenges in terms of storage, transmission, and deployment. 
Though great efforts have been devoted to model compression and quantization, existing methods often rely on fine-tuning or calibration data, which exhibit limited generalization across different tensor types.
In this paper, we argue that video codecs offer a promising solution for LLM compression, due to their inherent compatibility with matrix-structured data, configurable compression strategies, and the availability of highly optimized, off-the-shelf implementations.
Therefore, we present \textbf{LLMCodec}, a video codec-based LLM compression method that integrates affine quantization with the recent VVC/H.266 video codec. Beyond VVC, we further compare a range of video codecs and encoding profiles to evaluate their impact on compression performance.
Experiments on different models demonstrate the robustness and generality of LLMCodec. Notably, on LLaMA-3-8B at 2-bit precision, LLMCodec reduces perplexity by over 1.5× and improves downstream task accuracy by 21\% compared with the existing method.

\end{abstract}
\section{Introduction}
\label{sec:intro}

Large language models (LLMs) have demonstrated remarkable capabilities in tasks like dialogue, code generation, and information search. However, their increasing size brings high costs in training, storage, and deployment. As model parameters scale up, LLMs develop  emergent abilities, enabling them to handle more complex tasks—even those they were not explicitly trained for. This trend has led to the development of massive models. For instance, LLaMA-3-70B~\cite{llama3modelcard}, a widely used open-source model, requires approximately 140 GB of memory in FP16 format for inference and typically runs on at least two A100 GPUs (80 GB each) with tensor parallelism. 

To mitigate these demands, model compression techniques have been developed, with quantization being the most widely adopted. It reduces parameter bit-widths for weights, activations, and gradients, and can be applied via quantization-aware training (QAT), quantization-aware fine-tuning (QAF), or post-training quantization (PTQ), the latter being the most practical~\cite{survey-model-compression,modelcompressionzhou,modelquantization,ptq}. 
While these methods reduce model size with acceptable accuracy loss, they often depend on calibration data or fine-tuning, exhibiting limited generalization across different tensor types, potentially limiting their deployment flexibility.

\begin{figure}
    \centering

    \centering
    \includegraphics[width=0.8\linewidth]{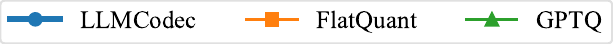}

    \begin{subfigure}{0.32\linewidth}
        \centering
        \includegraphics[width=\linewidth]{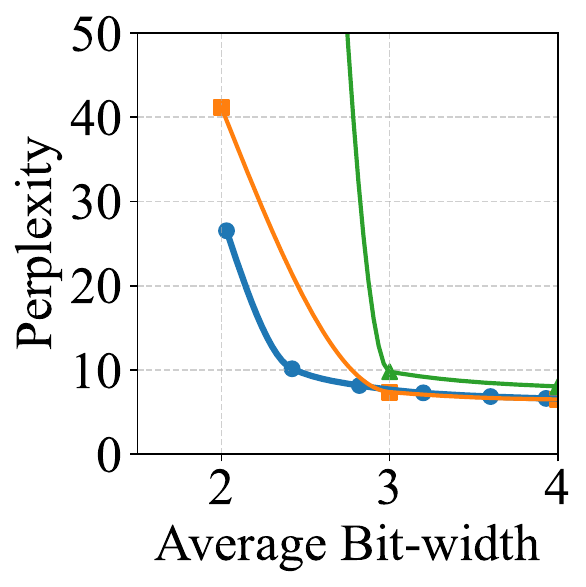}
        \caption{Wikitext2}
        \label{wikitext2}
    \end{subfigure}
    \hfill
    \begin{subfigure}{0.32\linewidth}
        \centering
        \includegraphics[width=\linewidth]{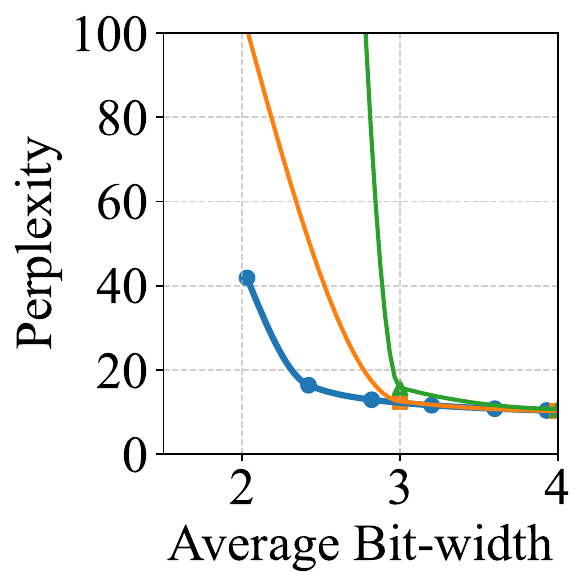}
        \caption{C4}
        \label{ARC-Challenge}
    \end{subfigure}
    \hfill
    \begin{subfigure}{0.32\linewidth}
        \centering
        \includegraphics[width=\linewidth]{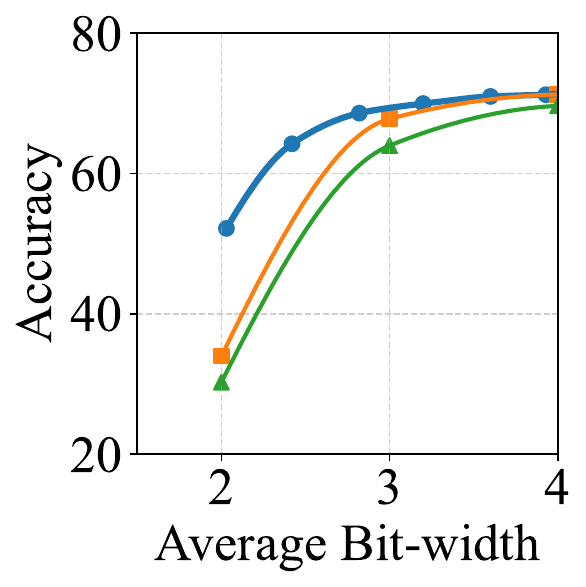}
        \caption{Tasks(Average)}
        \label{Hellaswag}
    \end{subfigure}
    
    \caption{Rate-distortion curves for LLaMA-3-8B compression: Comparison of LLMCodec, GPTQ~\cite{GPTQ}, and FlatQuant~\cite{flatquant} on PPL (WikiText2, C4) and average downstream task accuracy.}
    \label{fig:mainresults}
\end{figure}

In this paper, we argue that video codecs present a promising technique for LLM compression. First, video codecs are inherently well-suited for matrix-structured data, which aligns naturally with the format of neural networks' weights. Second, modern video codecs provide flexible compression configurations that allow fine-grained control of rate-distortion trade-offs, making them adaptable to different accuracy and storage requirements.

Therefore, we propose \textbf{LLMCodec}, a video codec-based LLM compression method. We integrate the outlier elimination strategy from FlatQuant~\cite{flatquant} into our framework, making parameters more suitable for codec compression. By systematically comparing different codecs and configurations, we identify effective strategies and compress LLMs to multiple sizes and performance levels.

Experimental results demonstrate that LLMCodec exhibits clear advantages under low-bit compression, outperforming baselines. Notably, when compressing the weights of LLaMA-3-8B~\cite{llama3modelcard} to 2 bits per parameter, it reduces perplexity on Wikitext2~\cite{wikitext} by approximately 36\% (from 41.15 to 26.53) and improves average downstream accuracy by 21\%. Results on other LLMs corroborate the effectiveness of LLMCodec in enhancing performance under extreme quantization. Our contributions are:

\begin{itemize}
    \item We propose LLMCodec, a video codec-based LLM compression framework that combines outlier elimination, RTN quantization, and modern video codecs.
    \item We extensively evaluate variant video codecs and encoding profiles to identify the optimal strategies for LLM compression.
    \item Experiments demonstrate that LLMCodec consistently improves quantization performance across multiple LLMs, in particular achieving up to a 36\% perplexity reduction on Wikitext2 and a 21\% accuracy improvement on downstream tasks for LLaMA-3-8B compared to FlatQuant.
\end{itemize}

\begin{figure*}[htbp]
    \centering
    \includegraphics[width=0.98\linewidth]{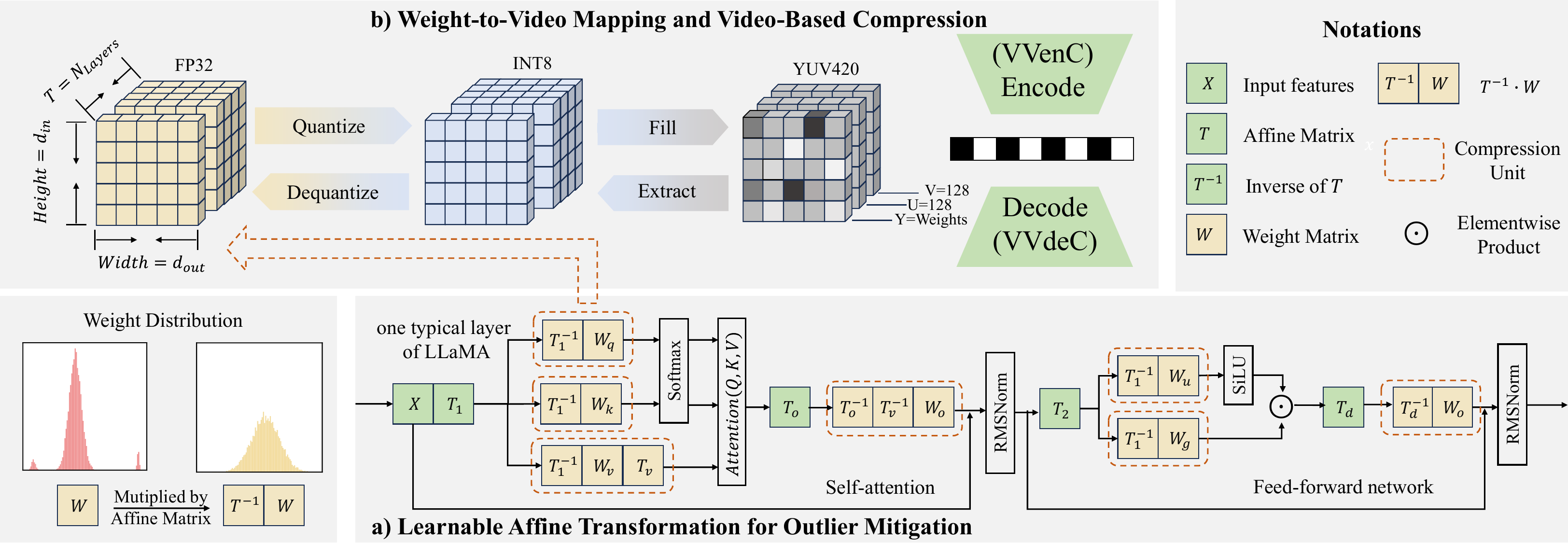}
    \caption{\textbf{Overview of LLMCodec.} The framework consists of two parts: \textbf{a) Learnable Affine Transformation for Outlier Mitigation:} applies affine transformations to eliminate weight outliers; \textbf{b) Weight-to-Video Mapping and Video-Based Compression:} converts tensors into video formats, applies compression/decompression with VVenC, and reconstructs them back into tensors.}
    \label{overview}
\end{figure*}

\section{Related work}
\label{sec:relate work}

\subsection{Model Quantization}

With the rapid growth of large language models, model compression has become an essential research direction. Among various techniques, quantization is the most widely adopted, as it reduces storage and computational costs by lowering the bit-width of parameter representations. Traditional per-layer or per-channel quantization methods, such as GPTQ~\cite{GPTQ}, employ gradient-based approximations to minimize quantization error. GPTQ achieves competitive performance without requiring full model fine-tuning and has become a representative approach in this domain. However, its performance degrades significantly under extremely low-bit settings (e.g., 2--3 bits).

Former studies have shown that model weights contribute unequally to overall performance, which has led to the development of mixed-precision quantization~\cite{llm.int8,SpQR,AWQ,BiLLM}. For instance, AWQ incorporates activation-aware strategies that assign higher precision to more important weights, improving downstream task performance. Despite these improvements, such methods involve high implementation complexity and remain limited at ultra-low bit-widths. Other approaches attempt to mitigate outlier weights through transformation techniques~\cite{quip,SmoothQuant,AffineQuant,flatquant}. The most recent example is FlatQuant~\cite{flatquant}, which applies affine transformations to weights and activations to reduce the impact of outliers, thereby achieving stronger results under low-bit quantization. Nevertheless, these methods often rely on retraining and lack flexible bit-width control or adaptive compression mechanisms.

Beyond centralized model compression, lightweighting techniques have also been explored in communication-constrained settings such as federated learning. For example, NPQ-MALA~\cite{npq-mala} combines pruning and quantization to reduce model size while maintaining accuracy. However, such approaches are typically designed for distributed training and are not optimized for static model deployment.

In summary, while existing quantization methods have made significant progress in reducing memory and computation, they still face challenges of bit-width rigidity, retraining overhead, and limited flexibility.

\subsection{Video Codecs}

Another line of related work originates from the field of signal and multimedia compression. Classical image and video coding standards, such as JPEG~\cite{jpeg}, WebP~\cite{webp}, HEVC/H.265~\cite{H265}, and VVC/H.266~\cite{H266}, are highly mature and widely deployed in industry. Their core principles---predictive coding, transform compression, and entropy coding---achieve high compression rates while preserving perceptual quality. More recently, neural networks have been integrated into compression pipelines, replacing traditional coding modules and enabling end-to-end audio-visual compression~\cite{Cheng2020LearnedIC}.

Recent studies further improve compression efficiency by exploiting rate--distortion optimization mechanisms inside standard codecs. For instance, latitude-guided and temporally dependent RDO strategies~\cite{latitude-rdo} adapt quantization parameters based on spatial and temporal characteristics, while CTU-level Lagrange multiplier adaptation~\cite{ctu-lagrange} enables finer-grained rate control in VVC. These works demonstrate the effectiveness of leveraging codec-internal optimization mechanisms to achieve better compression performance.

In parallel, several approaches combine learning-based pre- or post-processing with standard codecs to enhance compression efficiency. Rescaling-based frameworks~\cite{rescaling-framework} and context-guided resolution rescaling methods~\cite{cg-rrf} use neural networks to reshape signal distributions before encoding or to refine reconstruction quality after decoding, while remaining compatible with existing video coding standards.

These advances inspire a reverse perspective: instead of replacing traditional codecs with neural networks, mature video encoders can be repurposed for neural network model compression. Features intrinsic to video coding, such as rate--distortion controllability and robustness to distributional changes, are well aligned with the challenges of model compression.

Our work follows this direction by exploring the use of video encoding frameworks for large language model compression. By aligning model weight distributions with codec-friendly representations, our method addresses  limitations of existing quantization techniques, including fixed bit-widths, retraining cost, and limited generality.

\section{Method}
\label{sec:method}
\subsection{Overview}
\label{subsec:framework}

As shown in Fig.~\ref{overview}, our framework consists of two tightly coupled components: a learnable affine transformation for outlier mitigation and a video-based compression pipeline. 
These two components are jointly designed to reshape LLM weight distributions into a codec-friendly form and to efficiently exploit the rate–distortion optimization capabilities of modern video codecs.

\noindent\textbf{Learnable Affine Transformation for Outlier Mitigation.} 
Before compression, an affine transformation is applied to each weight matrix to suppress outliers that adversely affect both low-bit quantization and codec efficiency. The transformation matrix is learned through optimization to redistribute weight magnitudes. This process produces more homogeneous tensor representations with reduced dynamic range, which improves the effectiveness of video codec compression under low-bit constraints.

\noindent\textbf{Weight-to-Video Mapping and Video-Based Compression.} 
After outlier mitigation, the transformed weights are quantized from FP32 to INT8 using RTN (Round-to-Nearest) quantization, producing a codec-compatible integer representation. The quantized tensors are then reorganized into video sequences and mapped to the YUV420 format. The resulting sequences are compressed using a standard video codec. At deployment time, the compressed bitstream is decoded into quantized tensors, which are subsequently dequantized back to floating-point weights for inference.

This design combines the robustness of quantization with the efficiency of video compression. Compared with prior approaches that rely solely on quantization, LLMCodec achieves higher accuracy retention under the same compression ratio.

\subsection{Learnable Affine Transformation for Outlier Mitigation}
\label{subsec:Outlier Mitigation}

In low bit-width quantization, outliers in weight distributions often dominate the overall quantization error, leading to severe accuracy degradation. This issue is particularly pronounced in large language models, where a small fraction of extreme weight values can significantly distort the quantization scale and reduce the effective resolution for the majority of parameters. When weight tensors are mapped to image-like representations, extreme values manifest as a small number of pixels with very high intensity, while the majority of pixels remain clustered in a narrow low-amplitude range. This highly imbalanced distribution leads to poor signal homogeneity, making it difficult for video codecs to efficiently exploit spatial redundancy and predictability. As a result, mitigating the influence of outliers is a crucial prerequisite for achieving effective and robust compression when adopting video-codec-based representations under low bit-width constraints.

To address this challenge, we apply an invertible linear transformation to the weight matrix prior to quantization. The goal of this transformation is to redistribute the weight values more evenly across dimensions, thereby reducing the dynamic range disparity and alleviating the adverse impact of extreme values. Importantly, the transformation is designed to be fully invertible, ensuring that the original linear operation can be exactly recovered in full precision and only incurs approximation error after quantization. Our approach builds upon the recent FlatQuant~\cite{flatquant} method, which introduces a learnable affine transformation for each linear layer to minimize quantization-induced reconstruction error. Following this formulation, we optimize the affine transformation matrix by minimizing the discrepancy between the original and quantized layer outputs.

The optimization objective is formulated as:
\begin{equation}
    \mathbf{T}^{*}=\arg \min _{\mathbf{T}}\left\|\mathbf{Y}-\mathcal{Q}(\mathbf{X} \mathbf{T}) \,\mathcal{Q}\!\left(\mathbf{T}^{-1} \mathbf{W}\right)\right\|_{F}^{2},
\end{equation}

where $\mathbf{X}$ denotes the input features, $\mathbf{W}$ is the original weight matrix, and $\mathbf{Y}$ represents the output activations computed as $\mathbf{Y}=\mathbf{X}\mathbf{W}$. The matrix $\mathbf{T}$ is a learnable affine transformation applied to redistribute weight magnitudes, while $\mathcal{Q}(\cdot)$ denotes the quantization operator. The Frobenius norm $\|\cdot\|_{F}$ measures the reconstruction error between the original and quantized outputs.

By explicitly minimizing the output-level discrepancy, the learned transformation effectively suppresses extreme values and produces a weight distribution that is more suitable for subsequent codec-based compression.

\subsection{Weight-to-Video Mapping and Video-Based Compression}
\label{subsec:tensor2image}

In Transformer-based language models, weights are organized as two-dimensional tensors, which can be naturally represented as images. We map each weight tensor into the YUV420 format, which is the default input representation of most modern video codecs. In the YUV420 format, the luminance (Y) channel is encoded at full spatial resolution and higher precision, while the chrominance channels (U and V) are subsampled and encoded with lower precision. Since weight tensors from different layers should be represented with consistent precision, we store weight values exclusively in the Y channel, while the U and V channels are filled with constant values. Importantly, filling the chrominance channels with constants does not increase the bitrate. As these channels contain no spatial variation, they can be efficiently predicted and encoded with negligible cost by the codec. The effectiveness of this channel assignment strategy is further validated by experimental results in Section~\ref{subsec:coding-profiles}.

Since most video codecs operate on 8-bit integer inputs, the transformed weights are first quantized from FP32 to 8-bit integers using round-to-nearest (RTN) quantization:
\begin{equation}
\mathbf{W}_q = \text{round}\left(\frac{\mathbf{T}^{-1}\mathbf{W}}{s}\right),
\end{equation}
where $\mathbf{T}^{-1}\mathbf{W}$ denotes the FP32 weight matrix after outlier mitigation, $\mathbf{W}_q$ is the resulting quantized tensor, and $s$ is the scale factor that maps floating-point values into the 8-bit integer range. This step produces a codec-compatible representation while preserving the relative structure of the weights.

For compression, we adopt the VVC/H.266~\cite{H266} standard using the VVenC~\cite{vvenc} implementation. To ensure consistent and predictable compression behavior across layers, all tensors are encoded using the All-Intra profile. The overall compression ratio is controlled by adjusting the quantization parameter (QP), allowing flexible trade-offs between model size and inference accuracy. This design enables LLMCodec to support multiple target bit budgets using a unified compression pipeline. The impact of different coding profiles and codec configurations on compression efficiency and model performance is further analyzed in Section~\ref{subsec:coding-profiles}.

\section{Experiments}
\label{sec:experiments}
\subsection{Experimental Setup}

We conduct experiments on three representative models, LLaMA-3-8B~\cite{llama3modelcard}, LLaMA-2-7B~\cite{llama2}, and Qwen-2.5-Instruct-7B~\cite{qwen2.5}, evaluating perplexity on WikiText2~\cite{wikitext} and C4~\cite{c4}, as well as accuracy on several commonsense reasoning benchmarks~\cite{lm-eval}, including ARC-Challenge~\cite{arc}, HellaSwag~\cite{hellaswag}, LAMBADA~\cite{lambada}, PIQA~\cite{piqa}, and WinoGrande~\cite{winogrande}. For compression, we primarily employ the VVenC~\cite{vvenc} codec with All-Intra configuration, though other codecs such as JPEG~\cite{jpeg}, WebP~\cite{webp}, and H.265~\cite{H265} are included for comparison.  Experiments are performed on two NVIDIA RTX 4090 GPUs. We report model perplexity/accuracy and average bit-widths under varying compression rates. As baselines, we compare against GPTQ~\cite{GPTQ}, and FlatQuant~\cite{flatquant}.

\subsection{Compression Performance}

\begin{figure*}
    \centering

    \centering
    \includegraphics[width=0.4\linewidth]{figures/legend.pdf}

    \begin{subfigure}{0.16\linewidth}
        \centering
        \includegraphics[width=\linewidth]{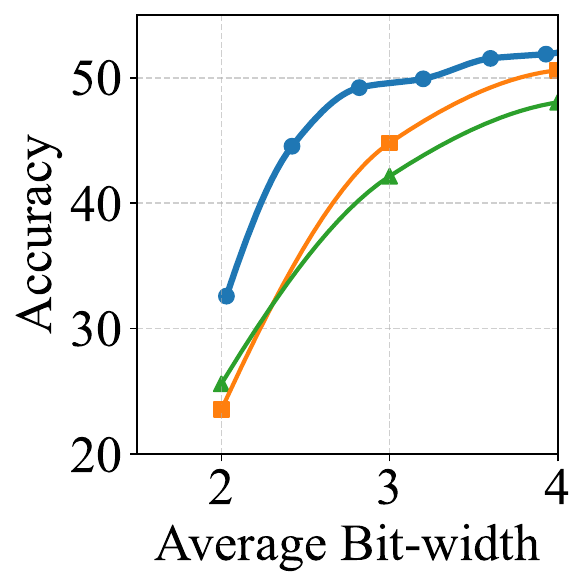}
        \caption{ARC-Challenge}
        \label{ARC-Challenge}
    \end{subfigure}
    \hfill
    \begin{subfigure}{0.16\linewidth}
        \centering
        \includegraphics[width=\linewidth]{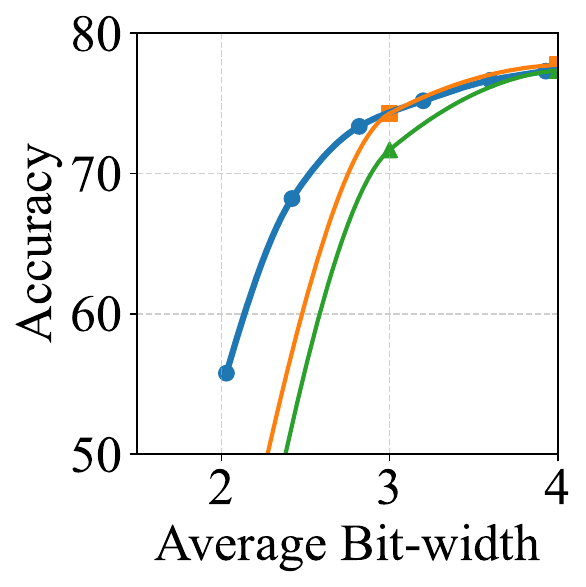}
        \caption{Hellaswag}
        \label{Hellaswag}
    \end{subfigure}
    \hfill
    \begin{subfigure}{0.16\linewidth}
        \centering
        \includegraphics[width=\linewidth]{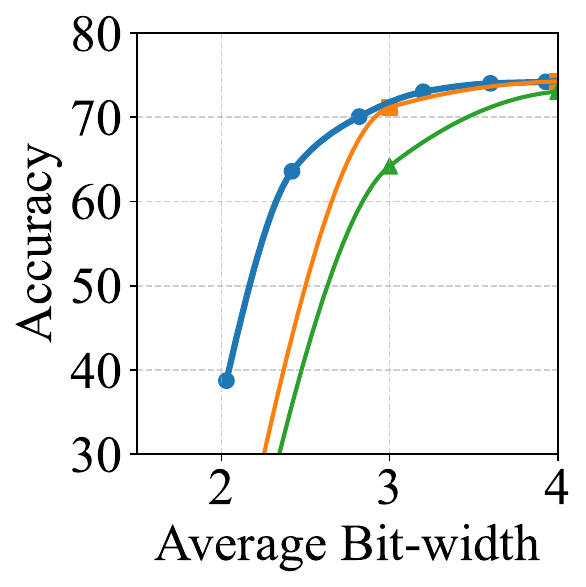}
        \caption{LAMBADA}
        \label{LAMBADA}
    \end{subfigure}
    \hfill
    \begin{subfigure}{0.16\linewidth}
        \centering
        \includegraphics[width=\linewidth]{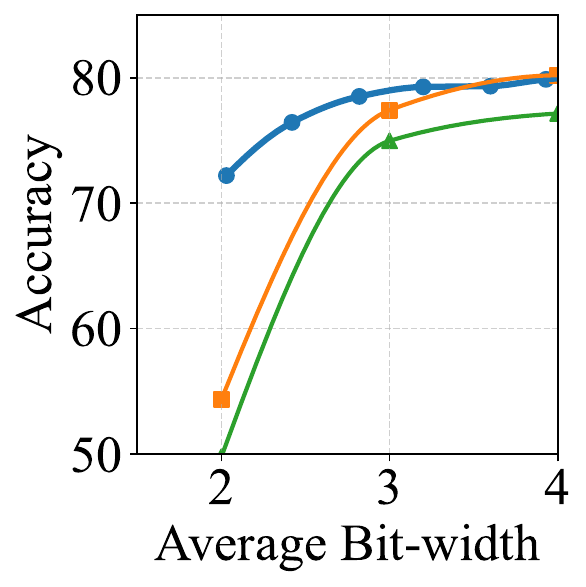}
        \caption{PIQA}
        \label{PIQA}
    \end{subfigure}
    \hfill
    \begin{subfigure}{0.16\linewidth}
        \centering
        \includegraphics[width=\linewidth]{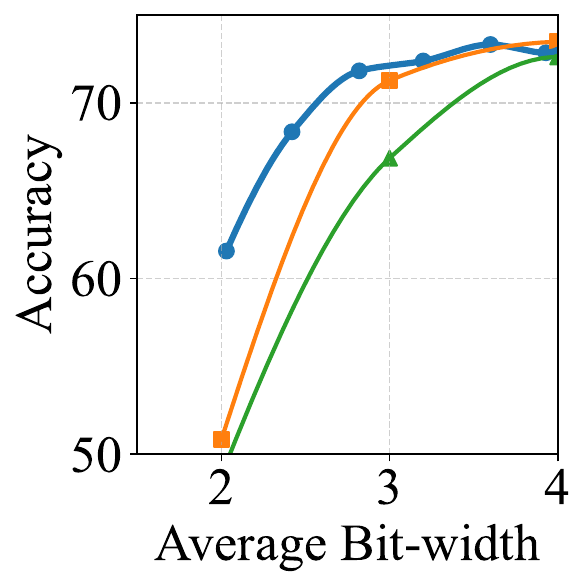}
        \caption{WinoGrande}
        \label{WinoGrande}
    \end{subfigure}
    \hfill
    \begin{subfigure}{0.16\linewidth}
        \centering
        \includegraphics[width=\linewidth]{figures/AvgAcc.pdf}
        \caption{Average}
        \label{wikitext2}
    \end{subfigure}

    \caption{Rate–distortion curves for LLaMA-3-8B compression evaluated on 5 downstream tasks and the average accuracy across all tasks. This figure presents a comparison of LLMCodec, GPTQ~\cite{GPTQ}, and FlatQuant~\cite{flatquant}.}
    \label{fig:mainresults-Accuracy}
\end{figure*}

We evaluate LLMCodec on LLaMA-3-8B, LLaMA-2-7B and Qwen-2.5-Instruct-7B across multiple compression rates and compare it with baselines. The results show that our method achieves performance comparable to prior approaches at high bit-width. However, as the bit-width decreases, LLMCodec consistently delivers lower perplexity and higher task accuracy, as summarized in Table~\ref{table:main-results} and further illustrated in Fig.~\ref{fig:mainresults} and Fig.~\ref{fig:mainresults-Accuracy}. Across different model families, LLMCodec consistently demonstrates  robustness under extreme compression. 
For the LLaMA-3-8B model at 2-bit precision, LLMCodec achieves a perplexity of 26.53 on WikiText2 and 41.86 on C4, significantly lower than FlatQuant’s 41.15 and 104.93, respectively. 
Moreover, LLMCodec attains a \textbf{21\% higher average accuracy} on downstream tasks, indicating better preservation of reasoning capabilities. 
On Qwen-2.5-7B-Instruct, our method also yields notable gains, reducing perplexity from 14.76 to 14.01 on WikiText2 and from 36.17 to 24.20 on C4, while improving downstream task accuracy by \textbf{16\%} on average. 
Although for LLaMA-2-7B the perplexity values are slightly higher than FlatQuant, LLMCodec still delivers a \textbf{10\% average accuracy improvement} across multiple tasks.

Overall, these results highlight LLMCodec’s ability to support flexible rate selection and to maintain strong performance in ultra-low-bit quantization settings, demonstrating clear advantages over existing baselines across diverse model families.

\begin{table*}[]
\caption{Compression performance of LLMCodec, FlatQuant~\cite{flatquant}, and GPTQ~\cite{GPTQ} on three LLMs under 3-bit and 2-bit weight quantization. The best results are highlighted in bold.}
\begin{center}
\begin{tabular}{ccccccccccc}
\Xhline{1.2pt}
\multirow{2}{*}{Model} & 
\multirow{2}{*}{\makecell{Average\\bit-width}} & 
\multirow{2}{*}{Method} & 
\multicolumn{2}{c}{Perplexity$\downarrow$} & 
\multicolumn{6}{c}{Accuracy$\uparrow$} \\ \cmidrule(lr){4-5} \cmidrule(lr){6-11}
 & & & Wikitext2 & C4 & ARC-Challenge & Hellaswag & LAMBADA & PIQA & WinoGrande & Average \\ \hline

 % \multicolumn{9}{c}{\textbf{LLaMA-3-8B}} \\ \hline
\multirow{7}{*}{\rotatebox{90}{LLaMA-3-8B}}
 & 16 & FP16 & 6.14 & 9.45 & 53.16 & 78.61 & 75.55 & 80.41 & 74.35 & 72.42 \\
 \cline{2-11}
 & \multirow{3}{*}{3} 
 & GPTQ~\cite{GPTQ}        & 9.79 & 15.79 & 42.15 & 71.68 & 64.20 & 74.97 & 66.85 & 63.97 \\
 & & FlatQuant~\cite{flatquant} & \textbf{7.37} & 12.54 & 44.80 & 74.27 & 71.18 & 77.42 & 71.27 & 67.79 \\
 & & \cellcolor{blue}\textbf{LLMCodec}       & \cellcolor{blue}7.61 & \cellcolor{blue}\textbf{12.23} & \cellcolor{blue}\textbf{50.77} & \cellcolor{blue}\textbf{74.38} & \cellcolor{blue}\textbf{72.15} & \cellcolor{blue}\textbf{78.78} & \cellcolor{blue}\textbf{73.40} & \cellcolor{blue}\textbf{69.90}\\ \cline{2-11}
 & \multirow{3}{*}{2} 
 & GPTQ~\cite{GPTQ}        & 465.23 & 1263.24 & 25.60 & 27.01 & 0.08 & 49.78 & 48.78 & 30.25 \\
 & & FlatQuant~\cite{flatquant} & 41.15 & 104.93 & 23.55 & 34.73 & 6.44 & 54.35 & 50.83 & 33.98 \\
 & & \cellcolor{blue}\textbf{LLMCodec}       & \cellcolor{blue}\textbf{26.53} & \cellcolor{blue}\textbf{41.86} & \cellcolor{blue}\textbf{32.59} & \cellcolor{blue}\textbf{55.77} & \cellcolor{blue}\textbf{38.73} & \cellcolor{blue}\textbf{72.20} & \cellcolor{blue}\textbf{61.56} & \cellcolor{blue}\textbf{52.17} \\
\hline

% \multicolumn{9}{c}{\textbf{Qwen-2.5-7B-Instruct}} \\ \hline
\multirow{7}{*}{\rotatebox{90}{\makecell{Qwen-2.5-7B\\-Instruct}}} &
16 & FP16 & 8.35 & 14.38 & 51.54 & 79.55 & 67.84 & 79.76 & 69.69 & 69.68 \\ 
\cline{2-11}
 & \multirow{3}{*}{3} 
 & GPTQ~\cite{GPTQ}        & 8.77 & 15.40 & 48.46 & 74.98 & 58.06 & 75.46 & 65.98 & 64.59 \\
 & & FlatQuant~\cite{flatquant} & \textbf{8.68} & 15.24 & 49.49 & 76.87 & \textbf{67.53} & \textbf{78.51} & \textbf{69.14} & 68.31 \\
 & & \cellcolor{blue}\textbf{LLMCodec}       & \cellcolor{blue}8.82 & \cellcolor{blue}\textbf{14.94} & \cellcolor{blue}\textbf{53.16} & \cellcolor{blue}\textbf{77.27} & \cellcolor{blue}66.27 & \cellcolor{blue}77.75 & \cellcolor{blue}68.82 & 
 \cellcolor{blue}\textbf{68.65} \\ \cline{2-11}
 &\multirow{3}{*}{2} 
 & GPTQ~\cite{GPTQ}        & 116.35 & 1196.20 & 26.54 & 27.88 & 0.1 & 49.51 & 49.88 & 30.78 \\
 & & FlatQuant~\cite{flatquant}& 14.76 & 36.17 & 31.14 & 51.31 & 25.77 & 60.88 & 53.67 & 44.55 \\
 & & \cellcolor{blue}\textbf{LLMCodec}       & \cellcolor{blue}\textbf{14.01} & \cellcolor{blue}\textbf{24.20} & \cellcolor{blue}\textbf{43.86} & \cellcolor{blue}\textbf{67.34} & \cellcolor{blue}\textbf{54.21} & \cellcolor{blue}\textbf{75.08} & \cellcolor{blue}\textbf{61.96} &
 \cellcolor{blue}\textbf{60.53} \\
\hline

% \multicolumn{9}{c}{\textbf{LLaMA-2-7B}} \\ \hline
\multirow{7}{*}{\rotatebox{90}{LLaMA-2-7B}} &
16 & FP16 & 5.47 & 7.26 & 46.25 & 75.99 & 73.88 & 79.11 & 69.06 & 68.86 \\ 
\cline{2-11}
 & \multirow{3}{*}{3} 
 & GPTQ~\cite{GPTQ}        & 6.34 & 9.07 & 40.61 & 71.00 & 66.89 & 76.28 & 67.25 & 64.41 \\
 & & FlatQuant~\cite{flatquant} & \textbf{6.05} & \textbf{7.86} & 40.87 & \textbf{72.18} & 71.16 & 77.31 & \textbf{67.88} & 65.88 \\
 & & \cellcolor{blue}\textbf{LLMCodec}       & \cellcolor{blue}6.33 & \cellcolor{blue}8.63 & \cellcolor{blue}\textbf{42.32} & \cellcolor{blue}71.54 & \cellcolor{blue}\textbf{71.16} & \cellcolor{blue}\textbf{78.13} & \cellcolor{blue}67.32 & \cellcolor{blue}\textbf{66.09} \\ \cline{2-11}
 & \multirow{3}{*}{2} 
 & GPTQ~\cite{GPTQ}        & 40.51 & 150.15 & 27.47 & 28.79 & 0.64 & 50.27 & 50.04 & 31.44 \\
 & & FlatQuant~\cite{flatquant} & \textbf{15.08} & \textbf{38.04} & 24.23 & 38.74 & 18.30 & 60.17 & 53.28 & 38.94 \\
 & & \cellcolor{blue}\textbf{LLMCodec}       & \cellcolor{blue}50.29 & \cellcolor{blue}72.69 & \cellcolor{blue}\textbf{28.58} & \cellcolor{blue}\textbf{50.14} & \cellcolor{blue}\textbf{37.8} & \cellcolor{blue}\textbf{68.88} & \cellcolor{blue}\textbf{59.12} & \cellcolor{blue}\textbf{48.90} \\
\Xhline{1.2pt}

\end{tabular}

\label{table:main-results}
\end{center}
\end{table*}

\subsection{Discussion on Video Codecs}
\label{subsec:coding-profiles}

We conduct ablation studies on video sequences reconstructed from the k\_projection layers of the LLaMA-3-8B model to identify the optimal coding profile for weight compression.

\begin{figure}
    \centering
        \begin{subfigure}{0.32\linewidth}
        \centering
        \includegraphics[width=\linewidth]{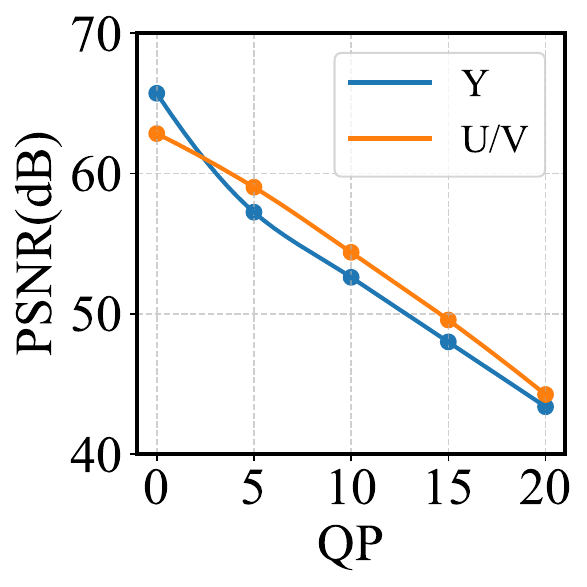}
        \caption{Per-Plane PSNR}
        \label{fig:yuv444}
    \end{subfigure} 
    \hfill
    \begin{subfigure}{0.32\linewidth}
        \centering
        \includegraphics[width=\linewidth]{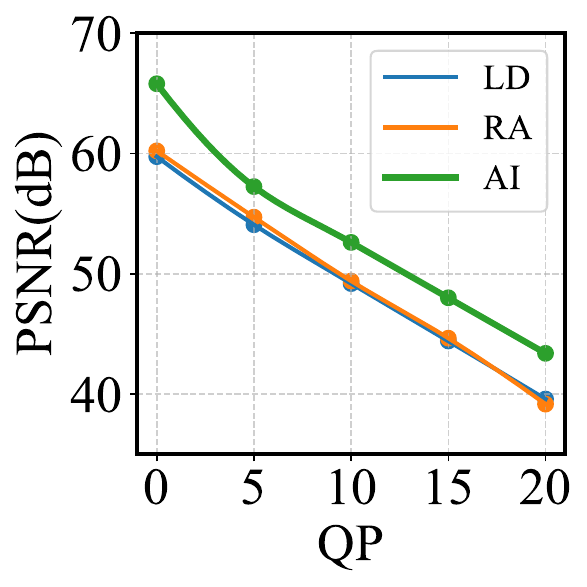}
        \caption{Coding Profiles}
        \label{fig:encoding-mode}
    \end{subfigure}
    \hfill
    \begin{subfigure}{0.32\linewidth}
        \centering
        \includegraphics[width=\linewidth]{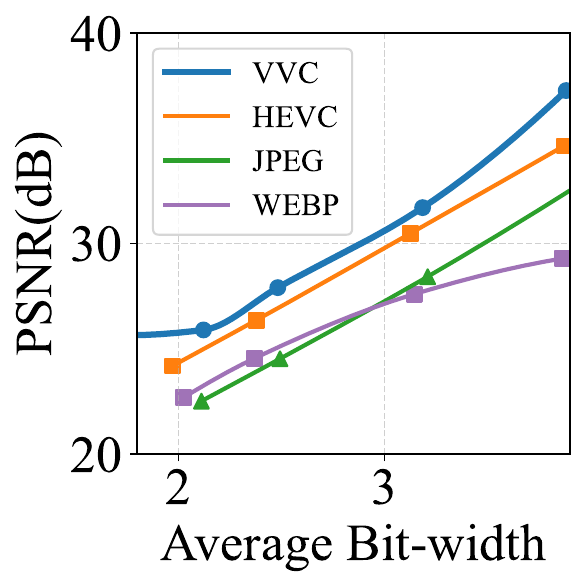}
        \caption{Video Codecs}
        \label{fig:codecs}
    \end{subfigure}

    \caption{R-D Curve of Different Settings}
    \label{coding profiles}
    
\end{figure}

\noindent\textbf{Padding Strategy.} We test padding model weights into all three YUV planes (i.e., YUV444) and plot the R–D curves for each. As shown in Fig.~\ref{fig:yuv444}, with identical QP settings and no offsets among planes, YUV444 still exhibits PSNR discrepancies across planes, leading to uneven precision among layers. To avoid this issue, we adopt the YUV420 format, where data are padded only into the Y plane.

\noindent\textbf{Different Coding Profiles.} We compare three profiles—All-Intra, Low-Delay, and Random-Access. As shown in Fig.~\ref{fig:encoding-mode}, All-Intra consistently achieves higher PSNR under the same QP setting. This suggests that the GOP-based prediction structure is not suitable for model weights. Different layers receive varying levels of precision under inter-frame coding, and model performance is ultimately constrained by the least accurate layer, resembling a “weakest-link” effect.

\noindent\textbf{Different Video Codecs.} Since LLM.265~\cite{vcllm} has found All-Intra profile is the best setting for weight compression, we further explore different image and video codecs, i.e., JPEG~\cite{jpeg}, WebP~\cite{webp}, HEVC/H.265~\cite{H265}, and VVC/H.266~\cite{H266}, and compare the compression performance for weights of LLMs.The rate–distortion results are presented in Fig.~\ref{fig:codecs}. As expected, the latest video coding standard, VVC/H.266, achieves the best overall performance, consistently preserving more model precision under the same compression rate.

\subsection{Ablation Studies}

To better understand the contribution of different preprocessing steps in our framework, we conducted ablation experiments on the LLaMA-3-8B model, comparing three configurations at an average bit-width of 2 bits:

\begin{itemize}
    \item \textbf{Without Affine Transformation}: raw weights are quantized to 8-bit with GPTQ and compressed without outlier removal.

    \item \textbf{Affine Transformation+GPTQ}: outliers are removed using learnable affine transformation,  followed by GPTQ quantization to 8-bit.
    
    \item \textbf{Affine Transformation+RTN (LLMCodec)}: outliers are removed using learnable affine transformation, followed by RTN quantization to 8-bit.
\end{itemize}

As shown in Table~\ref{ablation}, removing learnable affine transformation leads to a drastic degradation in performance, confirming that outliers dominate quantization error at extremely low bit-widths. When comparing quantization strategies, GPTQ provides no clear advantage over RTN in this setting. A possible reason is that GPTQ is optimized for direct low-bit quantization of weights, whereas our codec-based pipeline requires a uniform 8-bit intermediate representation. In this case, the lightweight RTN quantization proves sufficient when combined with video compression.

\begin{table}[htb]
\caption{Impact of Outlier Removal and Quantization Methods on LLMCodec at 2-bit Compression}
\begin{center}
\begin{tabular}{ccccc}
\hline
Method &
\makecell{Outlier\\Elimination} & 
\makecell{Weight\\Quantizer} & 
\makecell{Perplexity\\on Wikitext2} & 
\makecell{Average\\Accuracy} \\ \hline
\makecell{w/o affine trans.} & $\times$ & GPTQ & 1880.53 & 32.12 \\
\makecell{w/ GPTQ} & $\checkmark$ & GPTQ & 26.60 & 53.58 \\
\textbf{LLMCodec} & $\checkmark$ & RTN & 26.53 & 52.17 \\ \hline
\end{tabular}

\label{ablation}
\end{center}

\end{table}

\section{Conclusion}
\label{conclusion}

In this work, we proposed LLMCodec, a codec-based framework for compressing large language models. By combining outlier elimination, RTN quantization, and modern video codecs, LLMCodec aligns model weight distributions with codec compression mechanisms to achieve efficient compression. 
LLMCodec demonstrates robust performance improvements over FlatQuant~\cite{flatquant} across a range of language models, highlighting its generality under extreme quantization. These results highlight the potential of combining model quantization with mature video compression techniques to achieve efficient and scalable LLM deployment.

While this work focuses on static weight compression, several extensions remain open. Future efforts may explore activation compression to further reduce memory and bandwidth requirements, as well as integrating codec-aware operations into training or inference pipelines, further improving the practicality of codec-based compression for LLM deployment.

\section{Acknowledgment}
\label{Acknowledgement}

This work was partly supported by the NSFC(62431015, 62571317, 62501387), the Fundamental Research Funds for the Central Universities, Shanghai Key Laboratory of Digital Media Processing and Transmission under Grant 22DZ2229005, 111 project BP0719010.

% -------------------------------------------------------------------------
\bibliographystyle{IEEEbib}
\bibliography{refs}

\end{document}